\documentclass[namedreferences,hyperref,optionalrh]{spr-sola}
\usepackage{graphicx}        
\usepackage{color}           
\usepackage{breakurl}                         
\usepackage{longtable}                   

\chardef\us=`\_

\begin{document}

\begin{frontmatter}
\title{A Statistical Investigation of the Neupert Effect in Solar Flares Observed with ASO-S/HXI}

\author[addressref={aff1,aff2,aff3},email={lidong@pmo.ac.cn}]{\inits{D.}\fnm{Dong}~\snm{Li}\orcid{0000-0002-4538-9350}}
\author[addressref={aff1,aff2},email={ }]{\inits{ }\fnm{Hanyang}~\snm{Dong}\orcid{ }}
\author[addressref={aff1,aff2},email={ }]{\inits{ }\fnm{Wei}~\snm{Chen}\orcid{0000-0002-4118-9925}}
\author[addressref={aff1,aff2},email={ }]{\inits{ }\fnm{Yang}~\snm{Su}\orcid{0000-0002-4241-9921}}
\author[addressref={aff1,aff2},email={ }]{\inits{ }\fnm{Yu}~\snm{Huang}\orcid{0000-0002-0937-7221}}
\author[addressref={aff1,aff2},corref,email={ningzopngjun@pmo.ac.cn}]{\inits{Z.J.}\fnm{Zongjun}~\snm{Ning}\orcid{0000-0002-9893-4711}}
\address[id=aff1]{Key Laboratory of Dark Matter and Space Astronomy, Purple Mountain Observatory, CAS, Nanjing 210023, China}
\address[id=aff2]{School of Astronomy and Space Science, University of Science and Technology of China, Hefei 230026, China}
\address[id=aff3]{Yunnan Key Laboratory of the Solar physics and Space Science, Kunming 650216, PR China}

\runningauthor{Li et al.}
\runningtitle{The Neupert Effect in ASO-S/HXI Observed Flares}

\begin{abstract}
The Neupert effect refers to the strong correlation between the soft
X-ray (SXR) light curve and the time-integrated hard X-rays (HXR) or
microwave flux, which is frequently observed in solar flares. In
this article, we therefore utilized the newly launched {\it Hard X-ray
Imager} (HXI) on board the {\it Advanced Space-based Solar Observatory} to
investigate the Neupert effect during solar flares. By checking the
HXR light curves at 20$-$50~keV, a sample of 149 events that
cover the flare impulsive phase were selected. Then, we performed a
cross-correlation analysis between the HXR fluence (i.e., the time
integral of the HXR flux) and the SXR~1$-$8~{\AA} flux measured by
the {\it Geostationary Operational Environmental Satellite}. All the
selected flares show high correlation coefficients ($>$0.90),
which seem to be independent of the
flare location and class. The HXR fluences tend to increase linearly
with the SXR peak fluxes. Our observations indicate that
all the selected flares obey the Neupert effect.
\end{abstract}
\end{frontmatter}
\section{Introduction}
\cite{Neupert68} first noted that the shape of the soft X-ray (SXR) flux
appeared to resemble the time-integrated microwave light
curve during the impulsive phase of a solar flare. Such close
relationship was also observed between the SXR light curve and the
cumulative time integral of the hard X-ray (HXR) flux, this was dubbed as
`Neupert effect' \citep[e.g.][]{Hudson91,Yu21,Li24}. The empirical
correlation suggests that a causal relationship between the
nonthermal and thermal emissions could exist in a solar
flare. That is, the instantaneous radiation in the SXR waveband
depends on the accumulated energy that is deposited by the electron
beam during the flare impulsive phase
\citep[e.g.][]{Neupert68,Hudson91}. Based on the thick-target flare
model, the HXR emission is produced by the nonthermal bremsstrahlung
via Coulomb collisions, whereas the SXR emission is the cumulative
energy produced by the thermal bremsstrahlung that is also related
to the same electron beam. During this process, the heated plasma
via the thermal bremsstrahlung could rapidly expand up into the
upper atmosphere, called as `chromospheric evaporation'
\citep[e.g.][]{Fisher85,Battaglia09,Tian15,Tian18,Li15,Li22}.
Therefore, the Neupert effect could be regarded as an evidence of
the chromospheric evaporation driven by nonthermal electrons
\citep[e.g.][]{Lee95,McTiernan99,Mann06,Ning09,Ning10a,Li23}.

The Neupert effect has been studied by many authors since it was
reported by \cite{Neupert68}. Some authors estimated the nonthermal and thermal energies using X-ray spectroscopic imager observations to study the Neupert effect in solar flares \citep[e.g.][]{Datlowe74,Starr88,Feldman90,Dennis03,Ning08,Ning10b}.
This is because the SXR radiation is related to the thermal
energy accumulated in the solar corona, while the HXR emission is
linked to the nonthermal energy during the solar flare. However,
those observations had conflicting results. \cite{Dennis88} proposed
that the causal correlation between SXR and HXR emissions
could be attributed to different flare types or phases \citep[see
also][]{Dennis93}. The Neupert-type flare, which followed the Neupert effect, was usually found in the
model that the flare energy was released primarily by nonthermal
electrons \citep[e.g.][]{Brown71,Ning09}. Based on a statistical
investigation of the time difference between the SXR maximum and the end of the HXR emission in solar flares, \cite{Veronig02} concluded that the
chromospheric evaporation driven by electron beams plays a major
role in Neupert-type flares \citep[see
also][]{Veronig02b,Veronig05}. The Neupert effect was also observed in microflare data from the {\it Reuven
Ramaty High Energy Solar Spectroscopic Imager} (RHESSI) \citep[e.g.][]{Ning08b}.

The recently launched {\it Hard X-ray Imager} \citep[HXI;][]{Su19,Zhang19}
on board the {\it Advanced Space-based Solar Observatory}
\citep[ASO-S;][]{Gan19,Gan23,Huang19} provides a good chance to
study the relationship between the SXR light curve and the
time-integrated HXR flux during the impulsive phase of solar flares.
In this study, we statistically investigated the Neupert effect in
solar flares, which were simultaneously measured by ASO-S/HXI and
the {\it Geostationary Operational Environmental Satellite}
\citep[GOES;][]{Lotoaniu19}. The microwave emission of some flares
were measured by {\it Chashan Broadband Solar millimeter spectrometer}
\citep[CBS;][]{Shang22,Yan23} and the {\it Nobeyama Radio Polarimeters}
(NoRP), which were used to confirm the Neupert effect. The article
is organized as follows: Section~2 introduces the observations and
data analysis, Section~3 shows our primary results, Section~4
presents our discussion, and a brief summary is given in
Section~5.

\section{Observations and Data Analysis}
\subsection{Observations}
HXI is one of three payloads aboard ASO-S. It provides flare observations in the HXR energy range of $\approx$15$-$300~keV. The time
cadence is 4~s in the regular mode, and it can be as high as
0.125~s in the flare mode. Two types of data products are provided
by the HXI team, Level Q1 and Level 1. In our study, the
full-disk light curves at the energy channel of 20$-$50~keV are
used. The light curve is the sum of three flux monitors (D92, D93,
and D94), which are derived from the data production of Level Q1.
The time cadence of this data is 4~s in the regular mode
and becomes 2~s in the flare mode.

GOES is designed to continuously monitor the atmospheric conditions,
space weather, and solar activity. The {\it X-Ray Sensor} (XRS) aboard
GOES provides SXR light curves at 1$-$8~{\AA} and 0.5$-$4~{\AA},
which can be used for detecting solar flares \citep{Lotoaniu19}.
They have an uniform time cadence of 1~s. We also used the microwave
flux measured by CBS and NoRP to confirm the causal relationship
between the thermal and nonthermal emissions. CBS is a newly built
solar radio spectrograph at the Chashan Solar Observatory (CSO). It
is the first solar-dedicated radio spectrometer in the millimeter
regime of 35$-$40~GHz, which has a time cadence of $\approx$0.53~s
\citep{Shang23,Yan23}. NoRP provides the solar radio flux at six
frequencies, and it has a time cadence of 1~s.

\subsection{Data Analysis}
\begin{figure}
\centerline{\includegraphics[width=\textwidth,clip=]{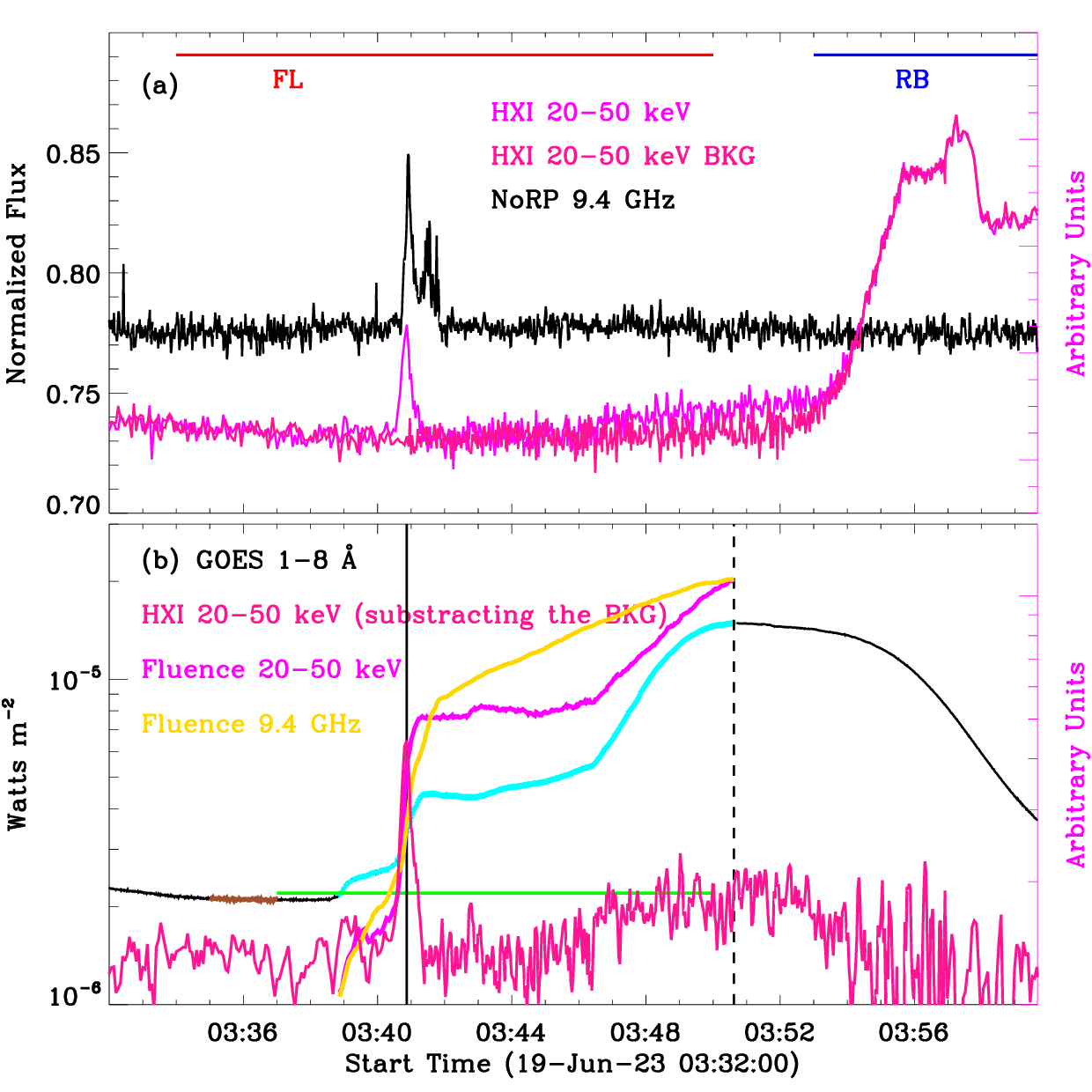}}
\caption{Overview of the solar flare on 19 Jun 2023. ({\bf a}): HXR and
microwave light curves measured by ASO-S/HXI (magenta and deep pink) and
NoRP (black), respectively. ({\bf b}): The SXR light curve recorded by
GOES (black), and the HXR flux (deep pink) after removing the background
emission. HXR and microwave fluence integrated over from HXI and
NoRP fluxes, respectively. The vertical lines mark the HXR peak
(solid) and the SXR maximum (dashed). The horizontal line
outlines the duration from the start and peak time given by the GOES
team.} \label{flux2}
\end{figure}

ASO-S/HXI has captured hundreds of solar flares since its first
light in the HXR energy range, which gives us an opportunity to
statistically study the HXR emission in solar flares. However, the HXI
data can be affected when ASO-S crosses the South Atlantic
Anomaly (SAA) and the radiation belt (RB), resulting in incomplete flare data. So, we first selected solar flares that
were observed completely by HXI, namely, the flares that showed significant enhancements during their impulsive phases at
HXI~20$-$50~keV, as shown in Figures~\ref{flux1} and \ref{flux2}.
The main reason for using HXI~20$-$50~keV flux is to
maintain a significant number of flare events. Moreover, the selected flares had to be recorded
by GOES at 1$-$8~{\AA}, because in this way we could study the flare
thermal emission in SXR. We finally
chose 149 flares from the one-year HXI observations, i.e. from 11
November 2022 to 11 November 2023 during the increasing period of
of Cycle 25, as listed in Table~\ref{tab1}. Here, there
are mainly two data biases in our event selection: (1) the 149
flares were completely recorded by GOES; (2) HXI~20$-$50~keV fluxes
showed obviously flare radiation during the impulsive phase.

In order to investigate the Neupert effect, we compared the cause
relationship between the HXR fluence ($\Gamma_{\rm HXR}$) and SXR
flux ($F_{\rm SXR}$) during the flare impulsive phase, as expressed
in previous articles \citep[e.g.][]{Lee95,Veronig02}

\begin{eqnarray}
\label{eq1}
     F_{\rm SXR}~\approx~k~\cdot~\Gamma_{\rm HXR}, \\
     \Gamma_{\rm HXR}~=~\int_{t_1}^{t_2}~(F_{\rm HXR}-F_{\rm BKG})~\cdot~dt.
\end{eqnarray}
\noindent where $F_{\rm HXR}$ and $F_{\rm BKG}$ denote the HXR light
curve and its background flux, which were observed by ASO-S/HXI at
20$-$50~keV. $t_1$ and $t_2$ are the start and end time of the HXR
light curve. The HXR fluence is defined as the time integral of the
HXR light curve after removing its background emission during the
impulsive phase of a solar flare. According to the theoretical
expectation that the HXR emission ends and the SXR maximum occurs at almost
the same time for the Neupert-type flare, the end time of the HXR
light curve is regarded as the maximum time of the SXR flux.

For a solar flare recorded by GOES, the start, peak, and stop time
is automatically provided by the GOES team. In this study, we
obtained these information from the Solar Flare Finder Widget, which
is available via the Solar Software \citep{Freeland98,Milligan18}.
This widget also provides the flare locations, which are calculated
from the {\it Atmospheric Imaging Assembly} images. Figure~\ref{flux1}b
presents the SXR flux recorded by GOES~1$-$8~{\AA}, the horizontal
green line outlines the duration of the impulsive phase;
the start ($t_s$) and peak ($t_p$) time is extracted from the Solar
Flare Finder Widget. The vertical lines mark the time when the HXR
(solid) and SXR (dashed) emissions reach their maximum,
respectively. Obviously, the SXR maximum time such as $t_2$ is
different from $t_p$ given by the GOES team; $t_2$ is
regarded as the end of the HXR emission. Similarly, $t_1$ is not equal to $t_s$. In
order to obtain $t_1$, the SXR emission with a duration of two
minutes before the solar flare ($t_s$) is first chosen as the
background, as marked by the brown pluses. Then, a threshold is
determined by the average value ($avg$) plus a statistically
significant (9~$\sigma$) of the background emission, as shown by the
green line. That is, $t_1$ occurs when the SXR flux exceeeds
the threshold ($avg+9~\sigma$). We want to state that $t_1$
and $t_2$ are identified from the SXR flux rather than the HXR light
curve, because the HXR light curve measured by HXI has a high
noise, especially for the non-flare flux. In such case, the
fluctuation of the background is too large, and it is impossible to
use a unique standard value. Figure~\ref{flux2}b shows another flare
recorded by GOES~1$-$8~{\AA}; similar temporal
properties confirm our selection.

\section{Results}
To show our statistical results, we first describe two events, as example.
\subsection{Examples in the Sample List}
\begin{figure}
\centerline{\includegraphics[width=\textwidth,clip=]{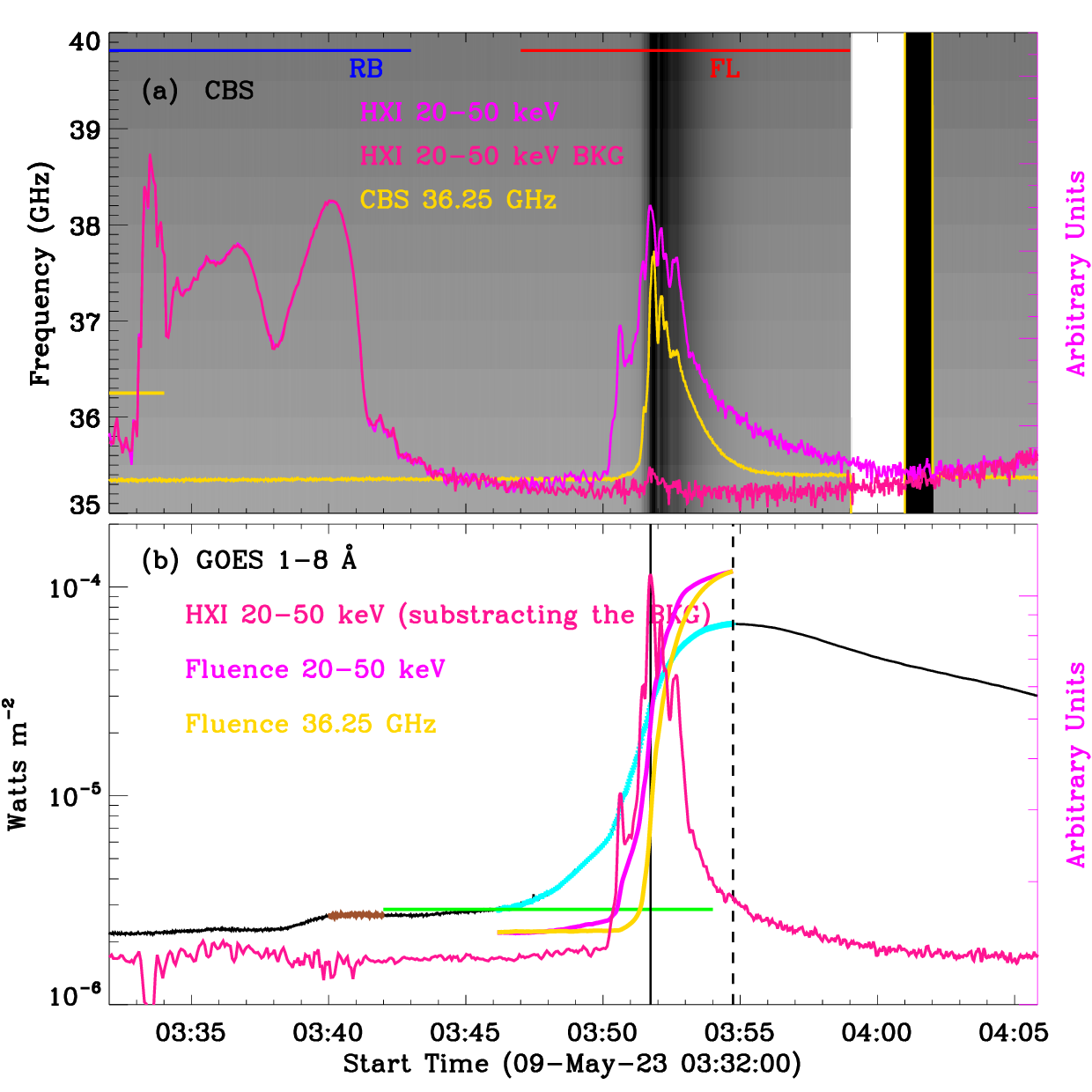}}
\caption{Overview of the solar flare on 09 May 2023. ({\bf a}): HXR and
microwave light curves measured by ASO-S/HXI (magenta and deep pink) and
CBS (gold), respectively. The context image is the radio dynamic
spectrum in the high frequency range of $\approx$35$-$40~GHz. ({\bf b}): The
SXR light curve recorded by GOES (black) and the HXR flux (deep pink)
after removing the background (BKG) emission. HXR and microwave fluences
integrated from HXI and CBS fluxes after removing their
background emissions. The vertical lines mark the HXR peak (solid)
and the SXR maximum (dashed). The horizontal line outlines the
duration from the start and peak time given by the GOES team.}
\label{flux1}
\end{figure}

Figure~\ref{flux1} shows an example of the solar flare that took place on
09 May 2023, which was an M6.5-class flare. Panel~a draws the HXR
light curves measured by HXI~20$-$50~keV, the magenta and deep pink
curves are the HXI data observed by three flux monitors and the
corresponding background monitors, respectively. It can be immediately noticed that the radiation enhancement in the approximate period of $\approx$03:33$-$03:44~UT
is from the background rather than from the solar flare, since the
magenta and deep pink curves overlap. This is mainly
because ASO-S goes through the radiation belt (RB) during
that time interval, as indicated by the blue line. On the other
hand, the radiation enhancement between about 03:45~UT and 03:55~UT
is the flare emission in the HXR channel and the background curve does
not reveal any significant enhancement. The microwave flux in the
frequency of 36.25~GHz (gold) further confirms that the
radiation is from a solar flare. The background image is the radio
dynamic spectrum in the  frequency range of 35$-$40~GHz observed by
CBS, which clearly shows a group of microwave bursts accompanying
the solar flare.

In Figure~\ref{flux1}b, we show the SXR flux recorded by GOES at
1$-$8~{\AA}, which starts at about 03:42~UT and peaks at around
03:54~UT\footnote{https://www.solarmonitor.org/?date=20230509}, as
outlined by the green line. We notice that both the SXR and HXR
(deep pink curve) fluxes do not exhibit a significant enhancement at about
03:42~UT, which is not suitable for the start time ($t_1$) of the
HXR fluence. Here, the time when the SXR flux exceeds the
threshold ($avg+9~\sigma$) is regarded as $t_1$, while the end
time ($t_2$) of the HXR fluence is the time when the SXR flux
reaches its maximum, based on the assumption that the SXR emission
reaches its maximum when the HXR ends \citep{Veronig02}. Then, the
HXR fluence can be calculated by integrating over the HXI flux after
removing the background emission between the time interval $t_1$ and
$t_2$, as shown by the magenta curve. It matches well the SXR
flux, as indicated by the cyan curve. Using the same method, the
microwave fluence is estimated by integrating over the CBS flux from
$t_1$ to $t_2$, as shown by the gold curve. It seems to be a bit
different from the SXR flux, which could be because the microwave
flux starts later than the HXR flux, as shown by the magenta and
gold line in Figure~\ref{flux1}a.

Figure~\ref{flux2} presents another solar flare that occurred on 19 June
2023, which was an M1.4-class flare. The HXR light curves were
observed by HXI~20$-$50~keV, and the microwave flux was detected by
NoRP~9.4~GHz. They both show a nonthermal pulse during the M1.4
flare, as shown in panel~a. The SXR flux recorded by
GOES~1$-$8~{\AA} suggests that the flare starts at about 03:37~UT
and peaks at around
03:50~UT\footnote{https://www.solarmonitor.org/?date=20230619}.
Similarly, these are not the HXR start and end times. Moreover, the SXR
flux reveals two primary peaks during the flare impulsive phase, as
shown in panel~b. Using the same method, the HXR/micorwave fluence
is estimated via integrating over the HXI and NoRP fluxes after
subtracting their background from $t_1$ to $t_2$, as indicated by
the magenta and gold curves. Those two curves appear to match
the SXR flux, particularly for the HXR fluence.

\begin{figure}
\centerline{\includegraphics[width=\textwidth,clip=]{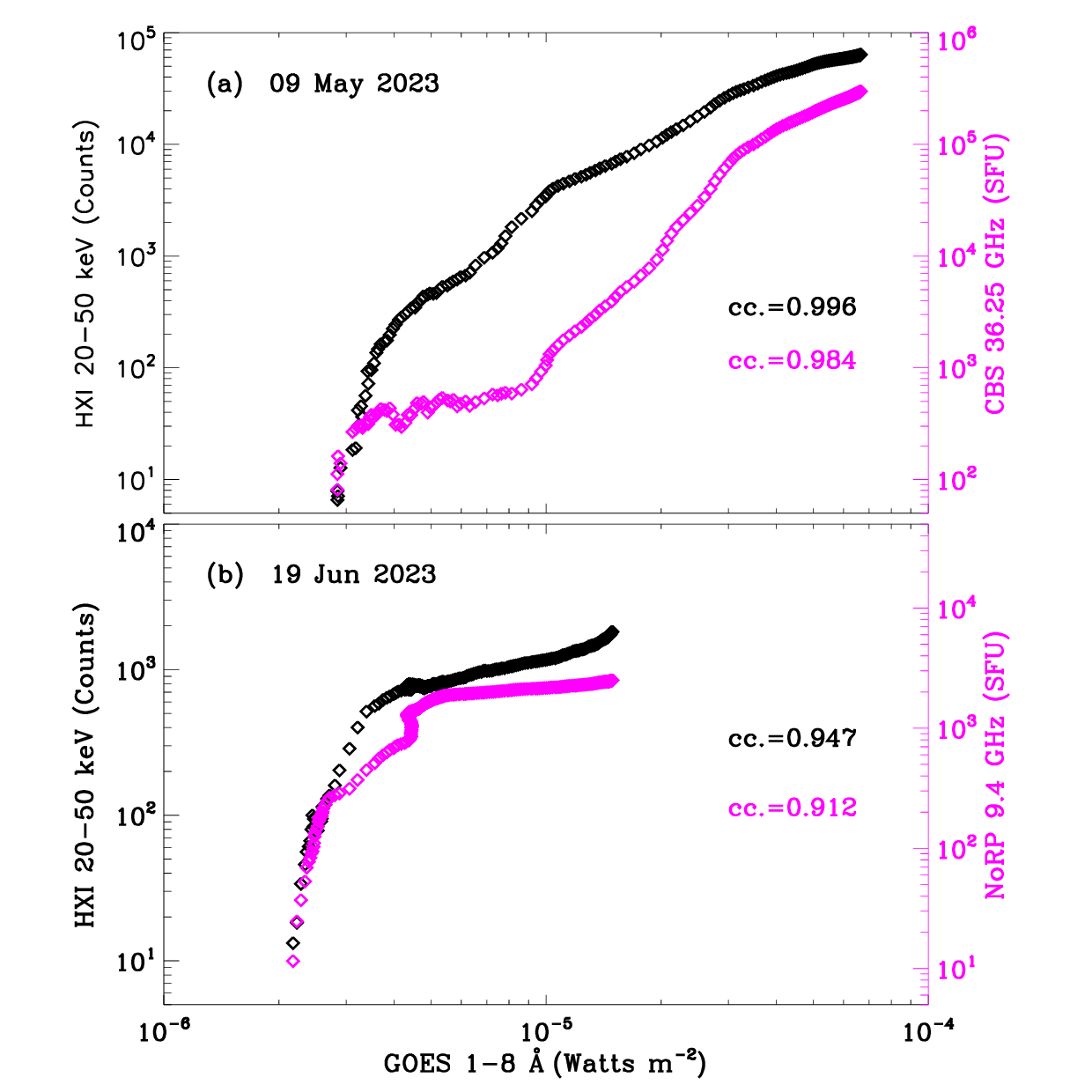}}
\caption{Correlation coefficients between the HXR and microwave fluences
and the SXR flux for the sampled flares on 09 May 2023 and 19 Jun
2023, respectively. The correlation coefficient values appear in the inset.}
\label{samp}
\end{figure}

To look closely at their cause relationship, we perform a
cross-correlation analysis between the SXR flux and the HXR and microwave fluences in the two flares, as shown in Figure~\ref{samp}. Here, the
time cadences of the SXR and microwave light curves are both
interpolated to the HXI time resolution, so they have the same
points during the same time interval. A very high correlation
coefficient (cc.) is obtained for the M6.5 flare on 09 May 2023,
implying that it is completely consistent with the Neupert effect.
Similarly to the observational result in Figure~\ref{flux1}b, the
correlation coefficient between the SXR flux and HXR fluence is
higher than that between the SXR flux and microwave fluence.
Conversely, the correlation coefficient for the M1.4 flare on 19
June 2023 is a little lower, which may suggest that it weakly agrees
with the Neupert effect.

\subsection{Statistical Results}
\begin{figure}
\centerline{\includegraphics[width=\textwidth,clip=]{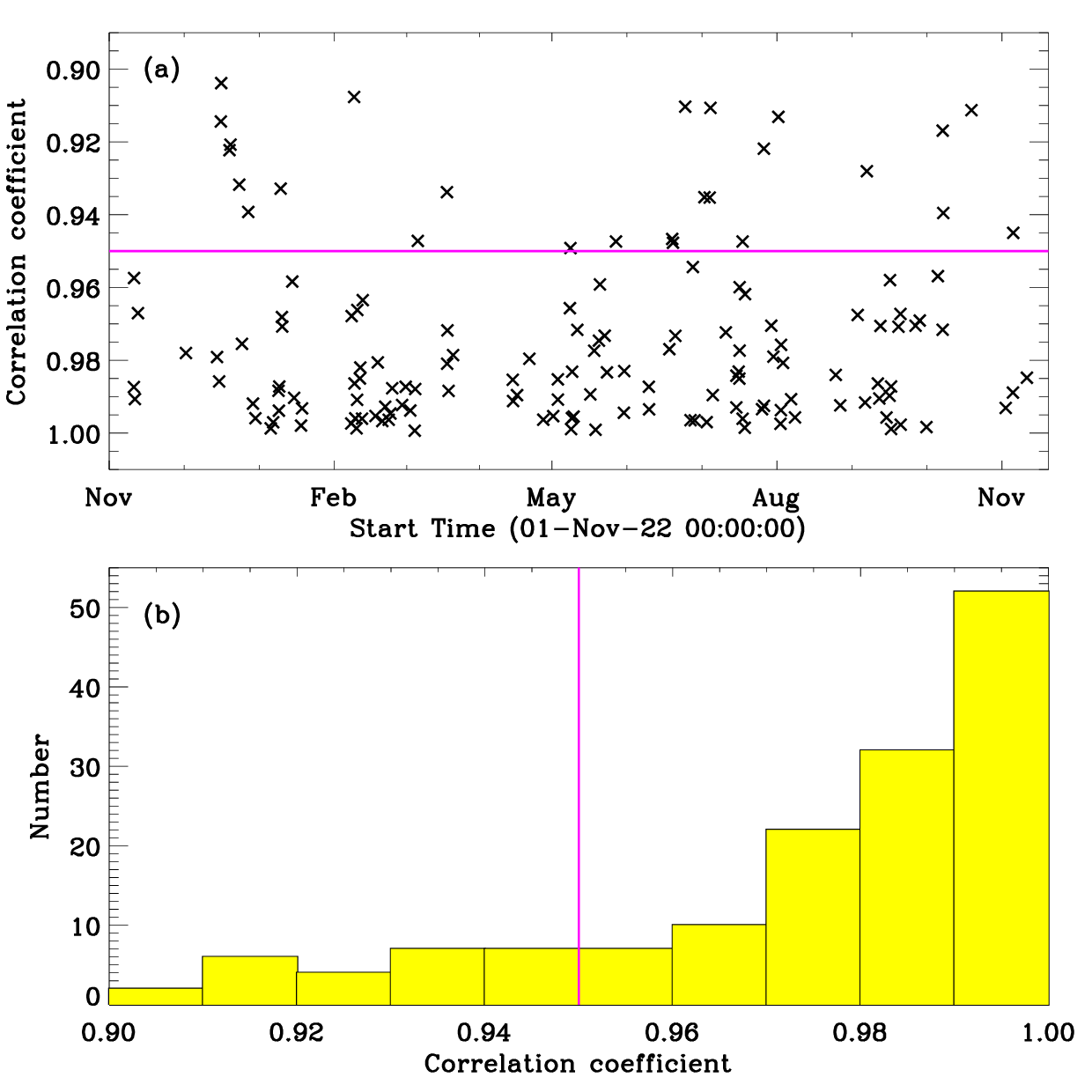}}
\caption{({\bf a}) Correlation coefficients varied as a function of time.
({\bf b}) Histogram of correlation coefficients. The magenta line marks the
correlation coefficient of 0.95.} \label{corr}
\end{figure}

Based on the same method, we investigated the simple correlation
between the HXR fluence and the SXR flux for 149 flares measured by
ASO-S/HXI and GOES. Figure~\ref{corr} presents the distribution of
correlation coefficients, namely, the correlation coefficients as a
function of time (panel a) and the histogram of the correlation coefficients
(panel b). We can see that all these solar flares show high correlation
coefficients, $>$0.90. Moreover, there are 123 flares that
show correlation coefficients above 0.95, and only 26 flares below 0.95, as shown in the histogram of the correlation coefficients. That is, about 82.5\% (123/149) of the studied flares show a strong correlation between the SXR flux and
the time integral of the HXR emission. Here, a threshold of 0.95
(magenta line) is chosen, since it is a middle value for the correlation
coefficients. We also considered the confidence interval at the
level of 95\% in wavelet analysis
\citep[e.g.][]{Scargle82,Torrence98}.

\begin{figure}
\centerline{\includegraphics[width=\textwidth,clip=]{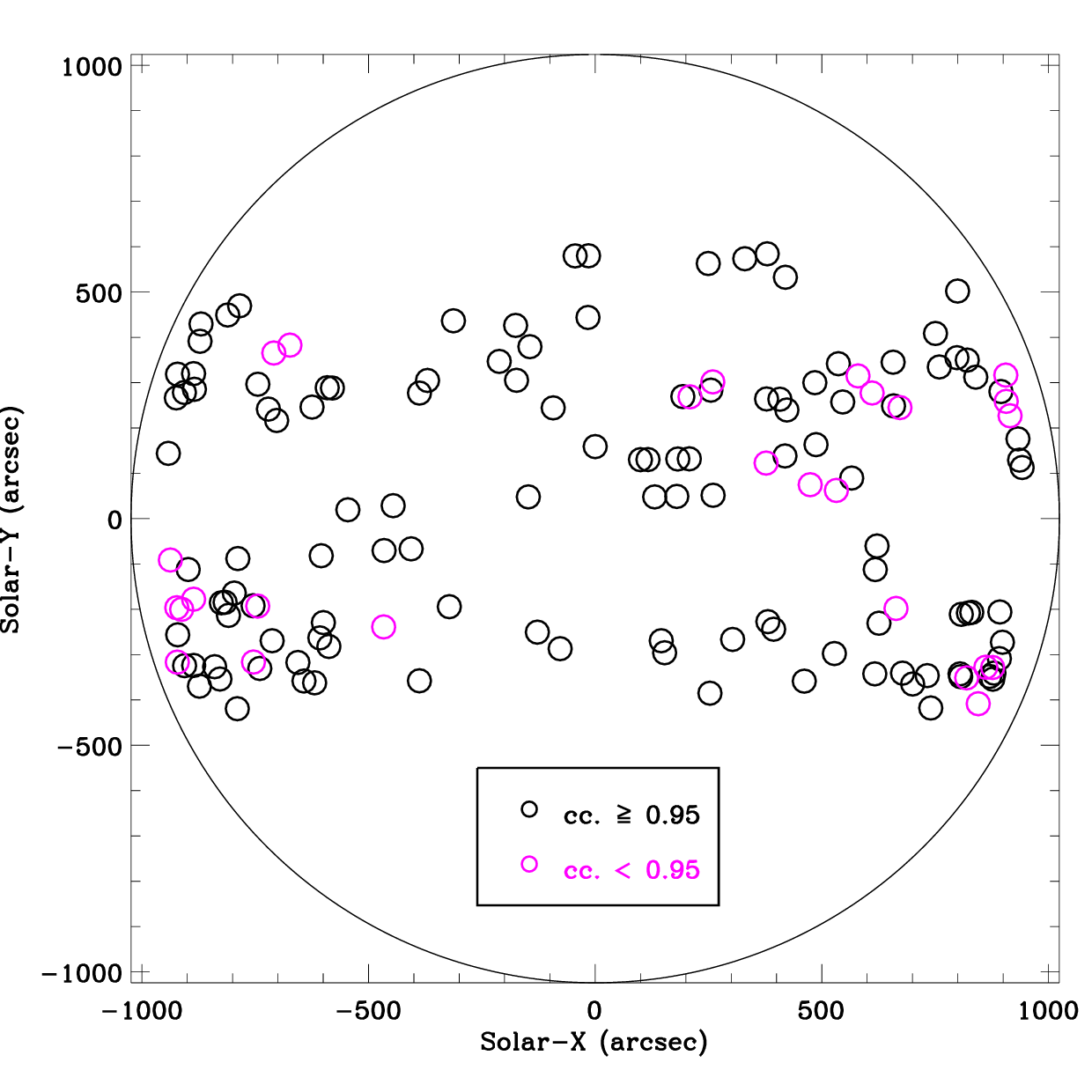}}
\caption{Positions of the studied flares on the Sun. The colors
represent the different ranges of correlation coefficients.}
\label{pos}
\end{figure}

Figure~\ref{pos} shows the locations of all these studied flares.
The colors represent the different ranges of the correlation
coefficients. As shown in Figure~\ref{pos}, these studied flares appear to be located
at the low and middle latitude, including all the
flares no matter if  they have higher ($\geq$0.95) or lower ($<$0.95) correlation coefficients. On the other hand, these flares that show lower
correlation coefficients seem to occur at the region that is close
to the solar limb, east or west solar limbs. While those
flares that have higher correlation coefficients appear in any
low- and middle-latitude region, i.e. the solar limb and solar center.

\begin{figure}
\centerline{\includegraphics[width=\textwidth,clip=]{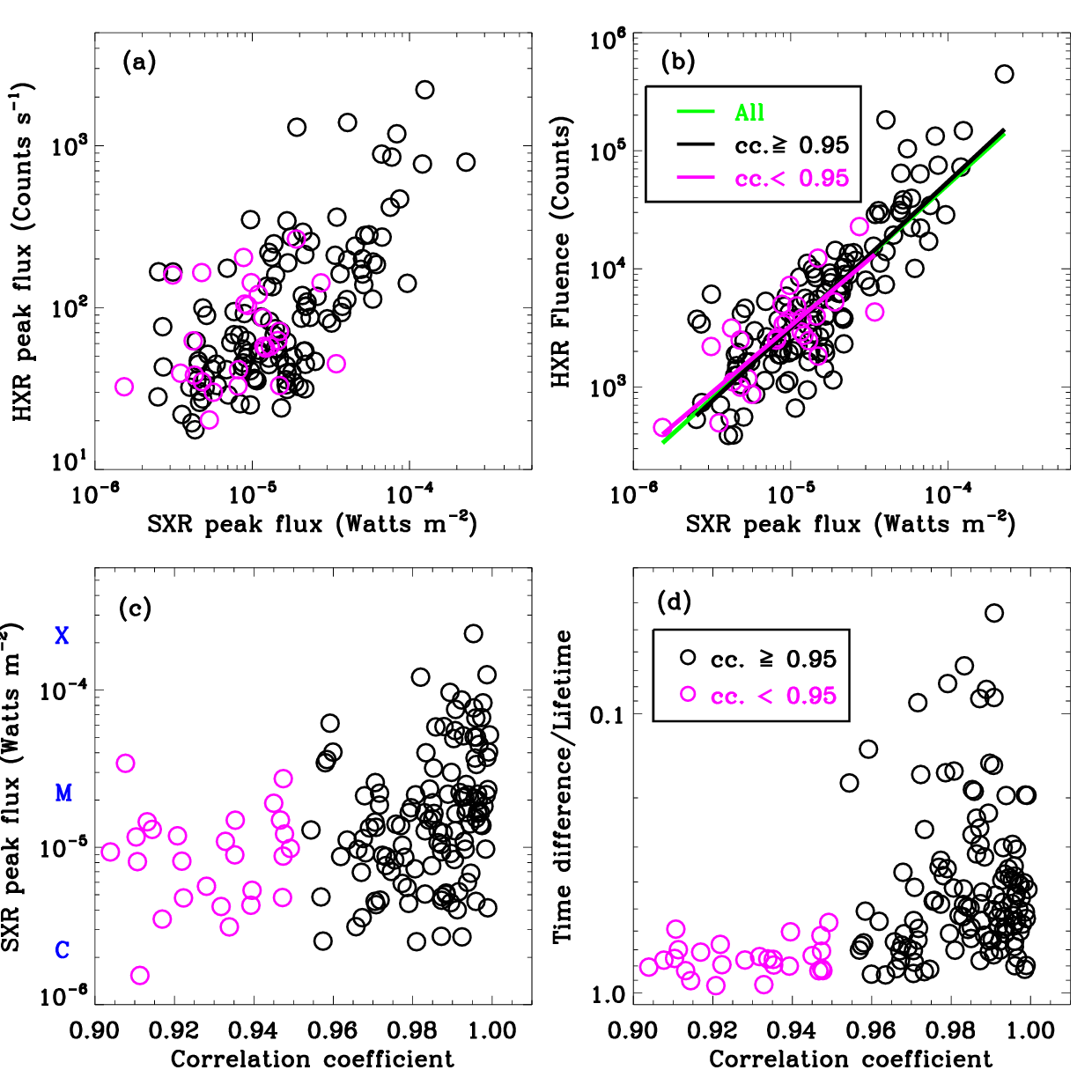}}
\caption{Scatter plots. ({\bf a}) HXR peak flux vs. SXR peak flux. ({\bf b}) HXR
fluence vs. SXR peak flux. The color lines indicate linear fitting
results. ({\bf c}) SXR peak flux  (or GOES class) vs. correlation
coefficient. ({\bf d}) The ratio between the time difference of the HXR peak and the SXR
maximum and the lifetime of the HXR fluence vs. the correlation coefficients.} \label{stat1}
\end{figure}

Figures~\ref{stat1} presents four scatter plots between some key
parameters. Panels~a and b show the HXR peak flux and fluence
depending on the SXR peak flux. Comparing with the HXR peak flux, the
HXR fluence exhibits a better linear relationship with the SXR peak
flux, as indicated by the black line in panel~b. Notice that the
statistical result is based on all the selected flares, whether they
have higher (cyan) or lower (magenta) correlation coefficients. The
linear fitting is also performed for those flares that have higher
and lower correlation coefficients, as indicated by the cyan and
magenta lines, respectively. The two fitting results (black and
magenta lines) are consistent with that for all the flares
(green line), which may suggest that all these flares are obeying
the Neupert effect. Figure~\ref{stat1}c displays the
SXR peak flux or GOES class depending on the correlation coefficient.
It seems that these lower correlation coefficients tend to be related to
small flares, while major flares tend to exhibit a higher
correlation. However, we should state that this is mainly because
the major flares are very rare in our study, i.e. only three
X-class flares. Panel~d presents the ratio between the time difference
and lifetime depending on the correlation coefficient. Here, the time
difference refers to the time delay between the HXR peak and the SXR
maximum, as marked by the vertical solid and dashed lines in
Figures~\ref{flux1}b and \ref{flux2}b. The lifetime refers to
the duration of the HXR fluence, i.e. $t_2-t_1$. The lower
correlation coefficients seem to appear in those flares that have a
large ratio between the time difference and the lifetime. For
instance, the ratio of those correlation coefficients below 0.95  is
close to 1.0.

\section{Discussion}
In this study, 149 events were selected to investigate the causal
relationship between the HXR fluence and the SXR flux, since the
impulsive phases of these flares were completely observed by
ASO-S/HXI in the energy range of 20$-$50~keV. The SXR fluxes were
recorded by GOES at 1$-$8~{\AA}, and the HXR fluence was defined as
the time integral of HXR light curves measured by HXI at
20$-$50~keV. The microwave fluxes, which were measured by CBS and
NoRP, were also used to examine the correlation between the
microwave fluence and the SXR flux. For those selected solar flares,
their HXR fluences were found to closely match the SXR fluxes,
implying that the studied flares could follow the Neupert effect.

In this work, the causal correlation between the time integral of the
HXR flux and the SXR emission is primarily identified as the
correlation coefficient between the HXR fluence and the SXR flux
during the flare impulsive phase. This is mainly because the
HXR emission produced by the nonthermal electron is strongly related
to the magnetic reconnection during the flare impulsive phase
\citep[e.g.][]{Krucker08,Shibata11,Jiang21,Li23b}. The duration of
the time integral is determined from the SXR flux observed by GOES,
since the HXR flux measured by ASO-S/HXI reveals a high noise level, which
could increase the uncertainty in the start and end times. The start
time ($t_1$) is regarded as the statistically significant excess of
9~$\sigma$ above the background emission, similarly to the triggered
mode of Konus-Wind\footnote{http://www.ioffe.ru/LEA/kwsun/}. The end
time ($t_2$) is defined as the maximum of the SXR flux,
according to the fact that the maximum of SXR light curves and the
end of HXR fluxes occur at nearly the same time \citep{Veronig02}.
Then, the lifetime of the HXR fluence can be obtained as
$t_2-t_1$. We also estimated the time difference between the peak
time of HXR pulse and the maximum time of the SXR flux. Noting that the
time difference studied here is not the same as that reported by
\cite{Veronig02}, who investigated the time delay between the SXR
maximum and the HXR end.

The correlation coefficients of all these 149 flares are greater
than 0.90, which may suggest a close relationship between the HXR
fluence and the SXR flux. The almost similar fitting results for these flares confirm the strong correlation. Moreover, about 82.5\% of the
studied flares show much higher correlation coefficients,
$\geq$0.95. The statistical results agree with the standard flare
model \citep[e.g.][]{Masuda94,Su13,Yan18,Yan22} or the CSHKP model
\citep[e.g.][]{Carmichael64,Sturrock66,Hirayama74,Kopp76},
indicating that the characterization we used in this study is reasonable.
Twenty six flares show lower correlation coefficients,
$<$0.95, but they are still larger than 0.90. All the lower
correlation coefficients tend to show a large ratio between the time
difference of the HXR peak and the SXR maximum and its lifetime ($t_2-t_1$).
For instance, their ratios could be as high as 1.0, indicating that
the acceleration process of nonthermal electrons is much shorter
than the heating process of thermal plasmas. Our observations
suggest that those solar flares might have an additional heating
process to the nonthermal electrons, which is consistent with the previous results \citep[e.g.][]{Veronig02,Li21,Yu21}.

For all these 149 flares, the correlation coefficients seem to be
independent on the flare location. They appear to take place in the
middle- and low-latitude regions and not in the
high-latitude region. While for these 26 flares that have lower
correlation coefficients, they appear to be distributed close to the
solar limb, i.e. the east and west limbs. The correlation coefficients do not exhibit a significant relation with the flare class or the
SXR peak flux. The lower correlation coefficients seem to appear in
small flares (C and M), which may because the intense flares
(X) are very rare in our study, only three X-class flares.
Similarly to pervious observations \citep[cf.][]{Veronig02}, the HXR
fluence displays a linear relationship with the SXR peak flux,
suggesting that the flare energy released during the impulsive phase
is primarily from the nonthermal electron beam
\citep{Priest02,Chen17,Warmuth16,Warmuth20}. Finally, we
want to state that the HXI and GOES data are selected during the increasing period of Solar Cycle 25, which might motivate the curiosity to perform a similar study for the later time of the solar cycle.

\section{Summary}
Using the HXR and SXR light curves observed by ASO-S/HXI and GOES,
we statistically investigated the Neupert effect during solar flares.
Our primary results are summarized as follows:

(1) 149 flares are selected from the HXI database. These flares
exhibit apparently radiation enhancements in the energy range of
20$-$50~keV, and their impulsive phases are completely recorded by
HXI.

(2) The selected flares show high correlation coefficients ($>$0.90) between the HXR fluence and the SXR flux. 123 (82.5\%) of
them exhibit much higher correlation coefficients, $\geq$0.95.

(3) The HXR fluence displays a linear relationship with the SXR peak
flux, and the correlation coefficient between the time integral of
the HXR flux and the SXR light curve appears to be independent of
the flare location and class. Our observational facts suggest that
all the selected flares could follow the Neupert effect

\begin{fundinginformation}
This work is funded by the National Key R\&D
Program of China 2022YFF0503002 (2022YFF0503000), NSFC under
grants 12073081, 12333010. D. Li is supported by Yunnan
Key Laboratory of Solar Physics and Space Science under the number
YNSPCC202207. This work is also supported by the Strategic Priority Research Program of
the Chinese Academy of Sciences, Grant No. XDB0560000. ASO-S mission is supported by the Strategic Priority Research Program on Space
Science, the Chinese Academy of Sciences, Grant No. XDA15320000.
\end{fundinginformation}

\begin{dataavailability}
ASO-S/HXI data: http://aso-s.pmo.ac.cn/sodc/dataArchive.jsp \\
GOES data: https://data.ngdc.noaa.gov/platforms/solar-space-observing-satellites/goes/ \\
NoRP data: http://solar.nro.nao.ac.jp/norp/index.html
\end{dataavailability}

\begin{authorcontribution}
Z.~J.~Ning provided the idea, D.~Li led this work and wrote the manuscript. H.~Y.~Dong participated in data analysis, W.~Chen, Y.~Su and Y.~Huang participated the HXI data correction and some discussions. All authors reviewed the manuscript.
\end{authorcontribution}

\begin{ethics}
\begin{conflict}
The authors declare that they have no conflicts of interest ..
\end{conflict}
\end{ethics}

\bibliographystyle{spr-mp-sola}
\bibliography{my_reference}

\clearpage

\clearpage
\renewcommand{\arraystretch}{1.4}
\begin{center}
\tabcolsep 3pt  
\small
\begin{longtable}{c c c c c c c c c}
\caption{Key parameters of the solar flares in this study.}
\label{tab1} \\
\endfirsthead
\multicolumn{7}{c}%
{{\bfseries \tablename\ \thetable{} -- Continued from previous page}} \\
\endhead
\hline \multicolumn{7}{l}%
{{Continued on next page}} \\
\endfoot
\hline \multicolumn{7}{l}%
{{$\bullet$ cc. is the correlation coefficient between the HXR fluence and the SXR flux.}} \\
\endlastfoot
\hline
No.  &  Date    &  \multicolumn{2}{c}{Time}  &  Flux (GOES)         & Fluence (HXI)   &  Position   &  $\bullet$ cc.   \\
     &          &    $t_1$     &    $t_2$    &  (W~m$^{-2}$)        & (Counts)        &   $X, Y$      &       \\
\hline
1 & 11 Nov 2022  & 01:31:14~UT  & 01:35:59~UT  &  4.85$\times$10$^{-6}$   & 4.15$\times$10$^3$  & 100$^{\prime\prime}$, 130$^{\prime\prime}$  & 0.987  \\
\hline
2 & 11 Nov 2022  & 03:15:28~UT  & 03:17:16~UT  &  2.54$\times$10$^{-6}$   & 3.74$\times$10$^3$  & 116$^{\prime\prime}$, 130$^{\prime\prime}$ & 0.957  \\
\hline
3 & 11 Nov 2022  & 11:33:13~UT  & 11:40:41~UT  &  1.28$\times$10$^{-5}$   & 1.11$\times$10$^4$  & 182$^{\prime\prime}$, 132$^{\prime\prime}$ & 0.991  \\
\hline
4 & 12 Nov 2022  & 18:01:55~UT  & 18:03:53~UT  &  6.94$\times$10$^{-6}$   & 5.31$\times$10$^3$  & 419$^{\prime\prime}$, 138$^{\prime\prime}$ & 0.967  \\
\hline
5 & 02 Dec 2022 & 09:15:20~UT  & 09:20:54~UT  &  8.65$\times$10$^{-6}$   & 1.86$\times$10$^3$  & -924$^{\prime\prime}$, 266$^{\prime\prime}$ & 0.978  \\
\hline
6 & 15 Dec 2022  & 01:29:03~UT  & 01:37:33~UT  &  1.67$\times$10$^{-5}$   & 1.94$\times$10$^3$  & 678$^{\prime\prime}$, -341$^{\prime\prime}$ & 0.979  \\
\hline
7 & 15 Dec 2022  & 22:23:52~UT  & 22:40:16~UT  &  5.84$\times$10$^{-5}$   & 2.24$\times$10$^4$  & 805$^{\prime\prime}$, -342$^{\prime\prime}$ & 0.986  \\
\hline
8 & 16 Dec 2022  & 15:35:11~UT  & 15:41:00~UT  &  1.30$\times$10$^{-5}$   & 2.47$\times$10$^3$  & 862$^{\prime\prime}$, -328$^{\prime\prime}$ & 0.914  \\
\hline
9 & 16 Dec 2022  & 19:04:49~UT  & 19:11:14~UT  &  9.35$\times$10$^{-6}$   & 3.53$\times$10$^3$  & 877$^{\prime\prime}$, -329$^{\prime\prime}$ & 0.904  \\
\hline
10 & 20 Dec 2022 & 04:09:19~UT  & 04:10:17~UT  &  4.77$\times$10$^{-6}$   & 2.50$\times$10$^3$  & -709$^{\prime\prime}$, 365$^{\prime\prime}$ & 0.922  \\
\hline
11 & 20 Dec 2022 & 13:52:46~UT  & 14:06:49~UT  &  1.18$\times$10$^{-5}$   & 3.81$\times$10$^3$  & -674$^{\prime\prime}$, 383$^{\prime\prime}$ & 0.921  \\
\hline
12 & 24 Dec 2022 & 04:09:41~UT  & 04:14:16~UT  &  4.21$\times$10$^{-6}$   & 3.15$\times$10$^3$  & 259$^{\prime\prime}$, 302$^{\prime\prime}$ & 0.932  \\
\hline
13 & 25 Dec 2022 & 06:56:45~UT  & 07:00:59~UT  &  8.32$\times$10$^{-6}$   & 3.70$\times$10$^3$  & 485$^{\prime\prime}$, 300$^{\prime\prime}$ & 0.975  \\
\hline
14 & 27 Dec 2022 & 20:35:16~UT  & 20:37:21~UT  &  4.28$\times$10$^{-6}$   & 1.10$\times$10$^3$  & -922$^{\prime\prime}$, -317$^{\prime\prime}$ & 0.940  \\
\hline
15 & 29 Dec 2022 & 18:15:04~UT  & 18:20:37~UT  &  2.23$\times$10$^{-5}$   & 1.14$\times$10$^4$  & -922$^{\prime\prime}$, 319$^{\prime\prime}$ & 0.992  \\
\hline
16 & 30 Dec 2022 & 19:08:04~UT  & 19:09:43~UT  &  4.53$\times$10$^{-6}$   & 1.51$\times$10$^3$  & -144$^{\prime\prime}$, 379$^{\prime\prime}$ & 0.996  \\
\hline
17 & 06 Jan 2023 & 00:47:58~UT  & 00:57:40~UT  &  1.25$\times$10$^{-4}$   & 1.48$\times$10$^5$  & -906$^{\prime\prime}$, -324$^{\prime\prime}$ & 0.999  \\
\hline
18 & 07 Jan 2023 & 00:44:25~UT  & 00:52:07~UT  &  1.66$\times$10$^{-5}$   &6.38$\times$10$^3$  & -840$^{\prime\prime}$, -327$^{\prime\prime}$ & 0.997  \\
\hline
19 & 09 Jan 2023 & 04:49:50~UT  & 04:51:18~UT  &  5.02$\times$10$^{-6}$   & 5.56$\times$10$^2$  & -921$^{\prime\prime}$, -256$^{\prime\prime}$ & 0.988  \\
\hline
20 & 09 Jan 2023 & 08:46:30~UT  & 09:01:18~UT  &  2.16$\times$10$^{-5}$   & 3.74$\times$10$^3$  & 381$^{\prime\prime}$, -227$^{\prime\prime}$ & 0.994  \\
\hline
21 & 09 Jan 2023 & 13:17:47~UT  & 13:22:14~UT  &  1.04$\times$10$^{-5}$   & 3.70$\times$10$^3$  & 394$^{\prime\prime}$, -244$^{\prime\prime}$ & 0.987  \\
\hline
22 & 10 Jan 2023 & 02:13:06~UT  & 02:16:18~UT  &  1.09$\times$10$^{-5}$   & 4.86$\times$10$^3$  & -886$^{\prime\prime}$, -178$^{\prime\prime}$ & 0.932  \\
\hline
23 & 10 Jan 2023 & 14:45:51~UT  & 14:47:36~UT  &  9.23$\times$10$^{-6}$   & 1.06$\times$10$^3$  & -825$^{\prime\prime}$, -186$^{\prime\prime}$ & 0.968  \\
\hline
24 & 10 Jan 2023 & 17:47:35~UT  & 17:48:31~UT  &  1.34$\times$10$^{-5}$   & 2.01$\times$10$^3$  & -816$^{\prime\prime}$, -185$^{\prime\prime}$ & 0.971  \\
\hline
25 & 14 Jan 2023 & 20:08:31~UT  & 20:21:22~UT  &  3.62$\times$10$^{-5}$   & 3.11$\times$10$^4$  & 626$^{\prime\prime}$, -231$^{\prime\prime}$ & 0.958  \\
\hline
26 & 15 Jan 2023 & 14:20:12~UT  & 14:31:34~UT  &  4.92$\times$10$^{-5}$   & 3.12$\times$10$^4$  & -809$^{\prime\prime}$, -214$^{\prime\prime}$ & 0.990  \\
\hline
27 & 18 Jan 2023 & 10:28:35~UT  & 10:34:58~UT  &  1.83$\times$10$^{-5}$   & 8.42$\times$10$^3$  & -322$^{\prime\prime}$, -194$^{\prime\prime}$ & 0.998  \\
\hline
28 & 18 Jan 2023 & 20:44:36~UT  & 20:47:23~UT  &  9.73$\times$10$^{-6}$   & 1.94$\times$10$^3$  & -212$^{\prime\prime}$, 347$^{\prime\prime}$ & 0.993  \\
\hline
29 & 07 Feb 2023 & 23:05:24~UT  & 23:07:36~UT  &  6.69$\times$10$^{-5}$   & 2.21$\times$10$^4$  & -44$^{\prime\prime}$, 580$^{\prime\prime}$ & 0.997  \\
\hline
30 & 08 Feb 2023 & 02:48:26~UT  & 02:53:40~UT  &  2.13$\times$10$^{-5}$   & 3.93$\times$10$^3$  & -15$^{\prime\prime}$, 580$^{\prime\prime}$ & 0.968  \\
\hline
31 & 09 Feb 2023 & 02:49:43~UT  & 03:10:37~UT  &  3.42$\times$10$^{-5}$   & 4.31$\times$10$^3$  & -937$^{\prime\prime}$, -91$^{\prime\prime}$ & 0.908  \\
\hline
32 & 09 Feb 2023 & 07:15:59~UT  & 07:17:59~UT  &  1.28$\times$10$^{-5}$   & 9.43$\times$10$^2$  & -898$^{\prime\prime}$, -112$^{\prime\prime}$ & 0.986  \\
\hline
33 & 09 Feb 2023 & 14:50:47~UT  & 14:56:46~UT  &  1.53$\times$10$^{-5}$   & 2.65$\times$10$^3$  & 250$^{\prime\prime}$, 563$^{\prime\prime}$ & 0.996  \\
\hline
34 & 10 Feb 2023 & 02:49:51~UT  & 03:03:55~UT  &  3.74$\times$10$^{-5}$   & 2.91$\times$10$^4$  & 330$^{\prime\prime}$, 573$^{\prime\prime}$ & 0.998  \\
\hline
35 & 10 Feb 2023 & 08:03:34~UT  & 08:05:44~UT  &  1.64$\times$10$^{-5}$   & 1.43$\times$10$^3$  & 379$^{\prime\prime}$, 584$^{\prime\prime}$ & 0.991  \\
\hline
36 & 10 Feb 2023 & 10:47:02~UT  & 10:49:21~UT  &  9.72$\times$10$^{-6}$   & 1.13$\times$10$^3$  & -788$^{\prime\prime}$, -88$^{\prime\prime}$ & 0.966  \\
\hline
37 & 11 Feb 2023 & 11:31:47~UT  & 11:34:36~UT  &  1.52$\times$10$^{-5}$   & 1.99$\times$10$^3$  & 800$^{\prime\prime}$, 502$^{\prime\prime}$ & 0.985  \\
\hline
38 & 11 Feb 2023 & 15:44:14~UT  & 15:48:45~UT  &  1.21$\times$10$^{-4}$   & 7.27$\times$10$^4$  & -604$^{\prime\prime}$, -82$^{\prime\prime}$ & 0.982  \\
\hline
39 & 12 Feb 2023 & 08:39:05~UT  & 08:48:04~UT  &  3.37$\times$10$^{-5}$   & 1.55$\times$10$^4$  & -466$^{\prime\prime}$, -70$^{\prime\prime}$ & 0.996  \\
\hline
40 & 12 Feb 2023 & 15:35:20~UT  & 15:37:57~UT  &  1.11$\times$10$^{-5}$   & 2.21$\times$10$^3$  & -406$^{\prime\prime}$, -67$^{\prime\prime}$ & 0.963  \\
\hline
41 & 17 Feb 2023 & 19:46:53~UT  & 20:16:54~UT  &  2.29$\times$10$^{-4}$   & 4.46$\times$10$^5$  & -811$^{\prime\prime}$, 449$^{\prime\prime}$ & 0.995  \\
\hline
42 & 18 Feb 2023 & 18:33:17~UT  & 18:35:52~UT  &  7.31$\times$10$^{-6}$   & 2.15$\times$10$^3$  & 423$^{\prime\prime}$, 239$^{\prime\prime}$ & 0.981  \\
\hline
43 & 20 Feb 2023 & 14:54:29~UT  & 14:58:15~UT  &  4.51$\times$10$^{-5}$   & 1.93$\times$10$^4$  & -870$^{\prime\prime}$, 429$^{\prime\prime}$ & 0.997  \\
\hline
44 & 21 Feb 2023 & 20:04:26~UT  & 20:16:59~UT  &  5.12$\times$10$^{-5}$   & 3.49$\times$10$^4$  & -785$^{\prime\prime}$, 469$^{\prime\prime}$ & 0.993  \\
\hline
45 & 23 Feb 2023 & 06:12:32~UT  & 06:13:55~UT  &  1.68$\times$10$^{-5}$   & 4.06$\times$10$^3$  & -313$^{\prime\prime}$, 437$^{\prime\prime}$ & 0.996  \\
\hline
46 & 23 Feb 2023 & 23:11:18~UT  & 23:13:38~UT  &  6.85$\times$10$^{-6}$   & 1.13$\times$10$^3$  & -176$^{\prime\prime}$, 426$^{\prime\prime}$ & 0.995  \\
\hline
47 & 24 Feb 2023 & 17:13:04~UT  & 17:15:41~UT  &  1.23$\times$10$^{-5}$   & 5.63$\times$10$^3$  & -16$^{\prime\prime}$, 444$^{\prime\prime}$ & 0.988  \\
\hline
48 & 28 Feb 2023 & 17:38:57~UT  & 17:50:28~UT  &  8.66$\times$10$^{-5}$   & 7.54$\times$10$^4$  & 420$^{\prime\prime}$, 532$^{\prime\prime}$ & 0.992  \\
\hline
49 & 02 Mar 2023 & 04:42:49~UT  & 04:50:17~UT  &  9.38$\times$10$^{-6}$   & 5.87$\times$10$^3$  & 751$^{\prime\prime}$, 409$^{\prime\prime}$ & 0.987  \\
\hline
50 & 04 Mar 2023 & 02:24:15~UT  & 02:25:56~UT  &  5.99$\times$10$^{-6}$   & 8.69$\times$10$^2$  & -703$^{\prime\prime}$, 217$^{\prime\prime}$ & 0.994  \\
\hline
51 & 05 Mar 2023 & 21:30:19~UT  & 21:35:47~UT  &  5.19$\times$10$^{-5}$   & 3.84$\times$10$^4$  & 798$^{\prime\prime}$, 355$^{\prime\prime}$ & 0.999  \\
\hline
52 & 06 Mar 2023 & 02:11:46~UT  & 02:28:13~UT  &  5.86$\times$10$^{-5}$   & 3.96$\times$10$^4$  & 821$^{\prime\prime}$, 350$^{\prime\prime}$ & 0.988  \\
\hline
53 & 07 Mar 2023 & 04:00:29~UT  & 04:05:01~UT  &  4.79$\times$10$^{-6}$   & 9.94$\times$10$^2$  & 906$^{\prime\prime}$, 317$^{\prime\prime}$ & 0.947  \\
\hline
54 & 19 Mar 2023 & 02:14:36~UT  & 02:15:42~UT  &  3.12$\times$10$^{-6}$   & 2.20$\times$10$^3$  & -754$^{\prime\prime}$, -316$^{\prime\prime}$ & 0.933  \\
\hline
55 & 19 Mar 2023 & 03:55:16~UT  & 03:56:38~UT  &  2.51$\times$10$^{-6}$   & 5.32$\times$10$^2$  & -740$^{\prime\prime}$, -331$^{\prime\prime}$ & 0.981  \\
\hline
56 & 19 Mar 2023 & 06:31:30~UT  & 06:35:31~UT  &  4.62$\times$10$^{-6}$   & 1.96$\times$10$^3$  & -740$^{\prime\prime}$, -331$^{\prime\prime}$ & 0.972  \\
\hline
57 & 19 Mar 2023 & 18:05:44~UT  & 18:08:48~UT  &  5.12$\times$10$^{-6}$   & 4.64$\times$10$^3$  & -656$^{\prime\prime}$, -317$^{\prime\prime}$ & 0.988  \\
\hline
58 & 21 Mar 2023 & 13:51:33~UT  & 13:53:57~UT  &  5.56$\times$10$^{-6}$   & 1.63$\times$10$^3$  & -619$^{\prime\prime}$, -362$^{\prime\prime}$ & 0.979  \\
\hline
59 & 14 Apr 2023 & 23:22:59~UT  & 23:27:51~UT  &  1.64$\times$10$^{-5}$   & 2.13$\times$10$^3$  & -721$^{\prime\prime}$, 242$^{\prime\prime}$ & 0.985  \\
\hline
60 & 15 Apr 2023 & 03:02:56~UT  & 03:03:26~UT  &  4.01$\times$10$^{-6}$   & 3.88$\times$10$^2$  & 618$^{\prime\prime}$, -112$^{\prime\prime}$ & 0.991  \\
\hline
61 & 16 Apr 2023 & 17:33:40~UT  & 17:43:46~UT  &  9.15$\times$10$^{-6}$   & 4.84$\times$10$^3$  & -872$^{\prime\prime}$, 391$^{\prime\prime}$ & 0.990  \\
\hline
62 & 21 Apr 2023 & 17:48:07~UT  & 18:11:44~UT  &  1.79$\times$10$^{-5}$   & 7.63$\times$10$^3$  & 153$^{\prime\prime}$, -296$^{\prime\prime}$ & 0.980  \\
\hline
63 & 27 Apr 2023 & 11:11:22~UT  & 11:13:53~UT  &  2.18$\times$10$^{-5}$   & 3.81$\times$10$^3$  & -77$^{\prime\prime}$, -287$^{\prime\prime}$ & 0.996  \\
\hline
64 & 01 May 2023 & 13:05:45~UT  & 13:09:25~UT  &  7.72$\times$10$^{-5}$   & 3.44$\times$10$^4$  & 617$^{\prime\prime}$, -343$^{\prime\prime}$ & 0.995  \\
\hline
65 & 03 May 2023 & 10:07:31~UT  & 10:14:16~UT  &  3.19$\times$10$^{-5}$   & 7.05$\times$10$^3$  & -625$^{\prime\prime}$, 246$^{\prime\prime}$ & 0.985  \\
\hline
66 & 03 May 2023 & 10:39:40~UT  & 10:45:20~UT  &  7.54$\times$10$^{-5}$   & 1.71$\times$10$^4$  & -625$^{\prime\prime}$, 246$^{\prime\prime}$ & 0.991  \\
\hline
67 & 08 May 2023 & 08:22:43~UT  & 08:25:48~UT  &  3.13$\times$10$^{-6}$   & 6.09$\times$10$^3$  & 193$^{\prime\prime}$, 269$^{\prime\prime}$ & 0.966  \\
\hline
68 & 08 May 2023 & 14:16:48~UT  & 14:21:03~UT  &  9.85$\times$10$^{-6}$   & 7.24$\times$10$^3$  & 209$^{\prime\prime}$, 268$^{\prime\prime}$ & 0.949  \\
\hline
69 & 08 May 2023 & 20:16:11~UT  & 20:25:19~UT  &  2.33$\times$10$^{-5}$   & 1.35$\times$10$^4$  & 255$^{\prime\prime}$, 283$^{\prime\prime}$ & 0.999  \\
\hline
70 & 09 May 2023 & 03:46:10~UT  & 03:54:44~UT  &  6.65$\times$10$^{-5}$   & 6.37$\times$10$^4$  & 407$^{\prime\prime}$, 263$^{\prime\prime}$ & 0.996  \\
\hline
71 & 09 May 2023 & 06:07:42~UT  & 06:13:32~UT  &  1.27$\times$10$^{-5}$   & 3.33$\times$10$^3$  & 378$^{\prime\prime}$, 264$^{\prime\prime}$ & 0.983  \\
\hline
72 & 09 May 2023 & 20:34:22~UT  & 20:52:38~UT  &  5.03$\times$10$^{-5}$   & 2.98$\times$10$^4$  & 546$^{\prime\prime}$, 257$^{\prime\prime}$ & 0.995  \\
\hline
73 & 11 May 2023 & 08:51:30~UT  & 09:01:38~UT  &  2.21$\times$10$^{-5}$   & 8.98$\times$10$^3$  & 622$^{\prime\prime}$, -60$^{\prime\prime}$ & 0.972  \\
\hline
74 & 16 May 2023 & 16:34:19~UT  & 16:43:26~UT  &  9.68$\times$10$^{-5}$   & 2.87$\times$10$^4$  & -873$^{\prime\prime}$, -370$^{\prime\prime}$ & 0.989  \\
\hline
75 & 18 May 2023 & 06:15:26~UT  & 06:26:14~UT  &  1.03$\times$10$^{-5}$   & 2.03$\times$10$^3$  & -906$^{\prime\prime}$, 278$^{\prime\prime}$ & 0.977  \\
\hline
76 & 18 May 2023 & 20:17:07~UT  & 20:23:36~UT  &  4.04$\times$10$^{-5}$   & 1.40$\times$10$^4$  & -884$^{\prime\prime}$, 285$^{\prime\prime}$ & 0.999  \\
\hline
77 & 20 May 2023 & 04:41:04~UT  & 04:50:06~UT  &  7.03$\times$10$^{-6}$   & 2.66$\times$10$^3$  & -790$^{\prime\prime}$, -419$^{\prime\prime}$ & 0.975  \\
\hline
78 & 20 May 2023 & 14:56:09~UT  & 14:59:48~UT  &  6.16$\times$10$^{-5}$   & 1.01$\times$10$^4$  & -744$^{\prime\prime}$, 297$^{\prime\prime}$ & 0.959  \\
\hline
79 & 22 May 2023 & 13:26:08~UT  & 13:27:28~UT  &  7.66$\times$10$^{-6}$   & 2.39$\times$10$^3$  & -370$^{\prime\prime}$, 305$^{\prime\prime}$ & 0.973  \\
\hline
80 & 23 May 2023 & 12:08:52~UT  & 12:13:53~UT  &  3.99$\times$10$^{-5}$   & 7.37$\times$10$^3$  & -174$^{\prime\prime}$, 305$^{\prime\prime}$ & 0.983  \\
\hline
81 & 27 May 2023 & 04:33:28~UT  & 04:35:42~UT  &  8.80$\times$10$^{-6}$   & 5.01$\times$10$^3$  & 611$^{\prime\prime}$, 277$^{\prime\prime}$ & 0.947  \\
\hline
82 & 30 May 2023 & 10:17:38~UT  & 10:21:58~UT  &  1.41$\times$10$^{-5}$   & 9.11$\times$10$^3$  & 877$^{\prime\prime}$, -354$^{\prime\prime}$ & 0.994  \\
\hline
83 & 30 May 2023 & 13:36:55~UT  & 13:38:53~UT  &  1.66$\times$10$^{-5}$   & 6.25$\times$10$^3$  & 528$^{\prime\prime}$, -298$^{\prime\prime}$ & 0.983  \\
\hline
84 & 09 Jun 2023 & 14:31:54~UT  & 14:40:00~UT  &  4.61$\times$10$^{-6}$   & 1.27$\times$10$^3$  & -128$^{\prime\prime}$, -250$^{\prime\prime}$ & 0.987  \\
\hline
85 & 09 Jun 2023 & 16:52:22~UT  & 17:11:26~UT  &  2.51$\times$10$^{-5}$   & 1.37$\times$10$^4$  & -642$^{\prime\prime}$, -358$^{\prime\prime}$ & 0.993  \\
\hline
86 & 18 Jun 2023 & 00:28:06~UT  & 00:31:37~UT  &  1.37$\times$10$^{-5}$   & 9.98$\times$10$^3$  & -388$^{\prime\prime}$, -358$^{\prime\prime}$ & 0.977  \\
\hline
87 & 19 Jun 2023 & 03:38:54~UT  & 03:50:37~UT  &  1.49$\times$10$^{-5}$   & 1.82$\times$10$^3$  & -923$^{\prime\prime}$, -197$^{\prime\prime}$ & 0.947  \\
\hline
88 & 19 Jun 2023 & 12:08:12~UT  & 12:14:18~UT  &  1.21$\times$10$^{-5}$   & 2.73$\times$10$^3$  & -912$^{\prime\prime}$, -200$^{\prime\prime}$ & 0.948  \\
\hline
89 & 20 Jun 2023 & 10:49:23~UT  & 11:06:05~UT  &  8.74$\times$10$^{-6}$   & 2.65$\times$10$^3$  & 806$^{\prime\prime}$, -349$^{\prime\prime}$ & 0.973  \\
\hline
90 & 24 Jun 2023 & 12:13:23~UT  & 12:17:03~UT  &  1.17$\times$10$^{-5}$   & 2.93$\times$10$^3$  & 580$^{\prime\prime}$, 315$^{\prime\prime}$ & 0.910   \\
\hline
91 & 26 Jun 2023 & 16:13:09~UT  & 16:22:22~UT  &  1.65$\times$10$^{-5}$   & 3.04$\times$10$^3$  & 536$^{\prime\prime}$, 342$^{\prime\prime}$ & 0.996  \\
\hline
92 & 27 Jun 2023 & 15:01:32~UT  & 15:14:25~UT  &  1.29$\times$10$^{-5}$   & 1.63$\times$10$^3$  & 657$^{\prime\prime}$, 345$^{\prime\prime}$ & 0.954  \\
\hline
93 & 28 Jun 2023 & 08:32:00~UT  & 08:44:47~UT  &  1.99$\times$10$^{-5}$   & 6.26$\times$10$^3$  & 759$^{\prime\prime}$, 335$^{\prime\prime}$ & 0.997  \\
\hline
94 & 02 Jul 2023 & 08:47:59~UT  & 08:52:50~UT  &  8.96$\times$10$^{-6}$   & 3.39$\times$10$^3$  & 672$^{\prime\prime}$, 245$^{\prime\prime}$ & 0.935  \\
\hline
95 & 03 Jul 2023 & 06:47:15~UT  & 06:53:14~UT  &  1.40$\times$10$^{-5}$   & 4.65$\times$10$^3$  & -588$^{\prime\prime}$, -282$^{\prime\prime}$ & 0.997  \\
\hline
96 & 04 Jul 2023 & 12:23:54~UT  & 12:35:53~UT  &  1.49$\times$10$^{-5}$   & 1.23$\times$10$^4$  & 907$^{\prime\prime}$, 259$^{\prime\prime}$ & 0.935  \\
\hline
97 & 04 Jul 2023 & 19:11:32~UT  & 19:27:51~UT  &  8.12$\times$10$^{-6}$   & 2.45$\times$10$^3$  & 915$^{\prime\prime}$, 227$^{\prime\prime}$ & 0.911  \\
\hline
98 & 05 Jul 2023 & 18:14:26~UT  & 18:17:05~UT  &  4.45$\times$10$^{-6}$   & 1.88$\times$10$^3$  & -886$^{\prime\prime}$, -324$^{\prime\prime}$ & 0.990  \\
\hline
99 & 11 Jul 2023 & 02:11:18~UT  & 02:14:13~UT  &  8.90$\times$10$^{-6}$   & 2.11$\times$10$^3$  & -886$^{\prime\prime}$, 320$^{\prime\prime}$ & 0.972  \\
\hline
100 & 15 Jul 2023 & 09:49:43~UT  & 09:53:15~UT  &  1.09$\times$10$^{-5}$   & 3.84$\times$10$^3$  & -591$^{\prime\prime}$, 289$^{\prime\prime}$ & 0.993  \\
\hline
101 & 15 Jul 2023 & 10:07:25~UT  & 10:09:53~UT  &  2.35$\times$10$^{-5}$   & 8.88$\times$10$^3$  & -580$^{\prime\prime}$, 288$^{\prime\prime}$ & 0.984  \\
\hline
102 & 16 Jul 2023 & 08:33:14~UT  & 08:40:48~UT  &  5.05$\times$10$^{-6}$   & 2.46$\times$10$^3$  & 732$^{\prime\prime}$, -346$^{\prime\prime}$ & 0.983  \\
\hline
103 & 16 Jul 2023 & 15:06:32~UT  & 15:09:02~UT  &  1.92$\times$10$^{-5}$   & 1.43$\times$10$^4$  & -388$^{\prime\prime}$, 277$^{\prime\prime}$ & 0.985  \\
\hline
104 & 16 Jul 2023 & 17:12:35~UT  & 17:19:24~UT  &  5.90$\times$10$^{-6}$   & 2.99$\times$10$^3$  & 700$^{\prime\prime}$, -365$^{\prime\prime}$ & 0.977  \\
\hline
105 & 16 Jul 2023 & 17:38:41~UT  & 17:48:13~UT  &  4.04$\times$10$^{-5}$   & 1.82$\times$10$^5$  & 740$^{\prime\prime}$, -418$^{\prime\prime}$ & 0.960  \\
\hline
106 & 17 Jul 2023 & 22:44:44~UT  & 22:55:20~UT  &  2.74$\times$10$^{-5}$   & 2.28$\times$10$^4$  & 845$^{\prime\prime}$, -408$^{\prime\prime}$ & 0.947  \\
\hline
107 & 17 Jul 2023 & 23:19:24~UT  & 23:34:30~UT  &  5.02$\times$10$^{-5}$   & 6.45$\times$10$^4$  & 870$^{\prime\prime}$, -348$^{\prime\prime}$ & 0.996  \\
\hline
108 & 18 Jul 2023 & 20:19:40~UT  & 20:28:07~UT  &  2.16$\times$10$^{-5}$   & 7.20$\times$10$^3$  & 880$^{\prime\prime}$, -341$^{\prime\prime}$ & 0.999  \\
\hline
109 & 18 Jul 2023 & 22:40:54~UT  & 22:44:12~UT  &  8.74$\times$10$^{-6}$   & 2.48$\times$10$^3$  & 892$^{\prime\prime}$, -309$^{\prime\prime}$ & 0.962  \\
\hline
110 & 25 Jul 2023 & 21:11:37~UT  & 21:16:47~UT  &  1.67$\times$10$^{-5}$   & 3.99$\times$10$^3$  & -600$^{\prime\prime}$, -229$^{\prime\prime}$ & 0.993  \\
\hline
111 & 26 Jul 2023 & 15:18:53~UT  & 15:24:03~UT  &  8.17$\times$10$^{-6}$   & 2.61$\times$10$^3$  & -467$^{\prime\prime}$, -239$^{\prime\prime}$ & 0.922  \\
\hline
112 & 26 Jul 2023 & 15:53:19~UT  & 15:59:41~UT  &  2.08$\times$10$^{-5}$   & 7.36$\times$10$^3$  & 840$^{\prime\prime}$, 312$^{\prime\prime}$ & 0.993  \\
\hline
113 & 29 Jul 2023 & 16:16:11~UT  & 16:24:14~UT  &  1.45$\times$10$^{-5}$   & 8.42$\times$10$^3$  & 146$^{\prime\prime}$, -269$^{\prime\prime}$ & 0.970  \\
\hline
114 & 30 Jul 2023 & 13:12:59~UT  & 13:15:51~UT  &  4.40$\times$10$^{-6}$   & 1.25$\times$10$^3$  & 303$^{\prime\prime}$, -267$^{\prime\prime}$ & 0.979  \\
\hline
115 & 01 Aug 2023 & 14:03:30~UT  & 14:09:16~UT  &  1.45$\times$10$^{-5}$   & 3.98$\times$10$^3$  & 664$^{\prime\prime}$, -198$^{\prime\prime}$ & 0.913  \\
\hline
116 & 02 Aug 2023 & 10:47:22~UT  & 10:49:48~UT  &  1.38$\times$10$^{-5}$   & 2.84$\times$10$^3$  & 808$^{\prime\prime}$, -211$^{\prime\prime}$ & 0.997  \\
\hline
117 & 02 Aug 2023 & 14:50:28~UT  & 14:52:27~UT  &  1.77$\times$10$^{-5}$   & 9.21$\times$10$^3$  & 823$^{\prime\prime}$, -208$^{\prime\prime}$ & 0.994  \\
\hline
118 & 02 Aug 2023 & 16:20:12~UT  & 16:22:19~UT  &  1.40$\times$10$^{-5}$   & 3.68$\times$10$^3$  & 831$^{\prime\prime}$, -207$^{\prime\prime}$ & 0.976  \\
\hline
119 & 03 Aug 2023 & 11:49:19~UT  & 11:55:10~UT  &  2.16$\times$10$^{-5}$   & 1.02$\times$10$^4$  & 893$^{\prime\prime}$, -206$^{\prime\prime}$ & 0.981  \\
\hline
120 & 06 Aug 2023 & 18:25:18~UT  & 18:40:32~UT  &  5.53$\times$10$^{-5}$   & 1.04$\times$10$^5$  & 936$^{\prime\prime}$, 129$^{\prime\prime}$ & 0.991  \\
\hline
121 & 08 Aug 2023 & 09:25:10~UT  & 09:31:16~UT  &  3.70$\times$10$^{-5}$   & 1.12$\times$10$^4$  & 895$^{\prime\prime}$, 280$^{\prime\prime}$ & 0.996  \\
\hline
122 & 25 Aug 2023 & 01:00:29~UT  & 01:09:15~UT  &  1.49$\times$10$^{-5}$   & 5.52$\times$10$^3$  & -713$^{\prime\prime}$, -269$^{\prime\prime}$ & 0.984  \\
\hline
123 & 26 Aug 2023 & 21:47:48~UT  & 21:50:20~UT  &  2.69$\times$10$^{-6}$   & 3.42$\times$10$^3$  & 942$^{\prime\prime}$, 113$^{\prime\prime}$ & 0.992  \\
\hline
124 & 03 Sep 2023 & 00:15:25~UT  & 00:23:00~UT  &  1.14$\times$10$^{-5}$   & 8.50$\times$10$^3$  & 933$^{\prime\prime}$, 176$^{\prime\prime}$ & 0.968  \\
\hline
125 & 06 Sep 2023 & 00:48:50~UT  & 00:50:29~UT  &  5.23$\times$10$^{-6}$   & 1.63$\times$10$^3$  & 208$^{\prime\prime}$, 132$^{\prime\prime}$ & 0.992  \\
\hline
126 & 06 Sep 2023 & 17:54:58~UT  & 17:56:29~UT  &  5.66$\times$10$^{-6}$   & 8.65$\times$10$^2$  & 377$^{\prime\prime}$, 122$^{\prime\prime}$ & 0.928  \\
\hline
127 & 11 Sep 2023 & 06:00:03~UT  & 06:01:16~UT  &  1.07$\times$10$^{-5}$   & 6.62$\times$10$^2$  & -796$^{\prime\prime}$, -164$^{\prime\prime}$ & 0.986  \\
\hline
128 & 11 Sep 2023 & 21:09:46~UT  & 21:14:02~UT  &  8.37$\times$10$^{-6}$   & 1.88$\times$10$^3$  & -93$^{\prime\prime}$, 244$^{\prime\prime}$ & 0.990  \\
\hline
129 & 12 Sep 2023 & 06:49:39~UT  & 07:02:15~UT  &  2.58$\times$10$^{-5}$   & 1.18$\times$10$^4$  & 487$^{\prime\prime}$, 163$^{\prime\prime}$ & 0.971  \\
\hline
130 & 14 Sep 2023 & 19:28:21~UT  & 19:31:52~UT  &  2.05$\times$10$^{-5}$   & 6.18$\times$10$^3$  & -148$^{\prime\prime}$, 48$^{\prime\prime}$ & 0.996  \\
\hline
131 & 16 Sep 2023 & 00:45:53~UT  & 00:49:49~UT  &  2.99$\times$10$^{-5}$   & 8.01$\times$10$^3$  & 131$^{\prime\prime}$, 48$^{\prime\prime}$ & 0.990  \\
\hline
132 & 16 Sep 2023 & 05:28:22~UT  & 05:38:45~UT  &  3.45$\times$10$^{-5}$   & 2.89$\times$10$^4$  & 180$^{\prime\prime}$, 49$^{\prime\prime}$ & 0.958  \\
\hline
133 & 16 Sep 2023 & 15:50:13~UT  & 15:51:00~UT  &  2.72$\times$10$^{-6}$   & 7.38$\times$10$^2$  & -942$^{\prime\prime}$, 144$^{\prime\prime}$ & 0.987  \\
\hline
134 & 16 Sep 2023 & 15:55:40~UT  & 15:56:57~UT  &  4.13$\times$10$^{-6}$   & 5.45$\times$10$^2$  & 260$^{\prime\prime}$, 52$^{\prime\prime}$ & 0.999  \\
\hline
135 & 19 Sep 2023 & 15:41:42~UT  & 15:52:32~UT  &  4.32$\times$10$^{-6}$   & 3.92$\times$10$^2$  & 658$^{\prime\prime}$, 249$^{\prime\prime}$ & 0.971  \\
\hline
136 & 20 Sep 2023 & 12:10:33~UT  & 12:12:49~UT  &  3.57$\times$10$^{-6}$   & 7.02$\times$10$^2$  & -446$^{\prime\prime}$, 29$^{\prime\prime}$ & 0.967  \\
\hline
137 & 20 Sep 2023 & 14:14:12~UT  & 14:19:06~UT  &  8.31$\times$10$^{-5}$   & 1.33$\times$10$^5$  & -546$^{\prime\prime}$, 20$^{\prime\prime}$ & 0.998  \\
\hline
138 & 26 Sep 2023 & 16:16:37~UT  & 16:18:08~UT  &  4.54$\times$10$^{-6}$   & 1.42$\times$10$^3$  & 253$^{\prime\prime}$, -385$^{\prime\prime}$ & 0.971  \\
\hline
139 & 28 Sep 2023 & 09:05:38~UT  & 09:07:30~UT  &  1.33$\times$10$^{-5}$   & 5.21$\times$10$^3$  & -828$^{\prime\prime}$, -354$^{\prime\prime}$ & 0.969  \\
\hline
140 & 01 Oct 2023 & 03:22:27~UT  & 03:24:07~UT  &  9.75$\times$10$^{-6}$   & 4.25$\times$10$^3$  & 898$^{\prime\prime}$, -273$^{\prime\prime}$ & 0.998  \\
\hline
141 & 05 Oct 2023 & 19:09:32~UT  & 19:13:25~UT  &  4.84$\times$10$^{-6}$   & 1.05$\times$10$^3$  & 0$^{\prime\prime}$, 159$^{\prime\prime}$ & 0.957  \\
\hline
142 & 07 Oct 2023 & 18:03:04~UT  & 18:06:19~UT  &  1.86$\times$10$^{-5}$   & 1.14$\times$10$^3$  & -607$^{\prime\prime}$, -263$^{\prime\prime}$ & 0.972  \\
\hline
143 & 07 Oct 2023 & 18:27:57~UT  & 18:29:59~UT  &  3.50$\times$10$^{-6}$   & 4.97$\times$10$^2$  & 474$^{\prime\prime}$, 75$^{\prime\prime}$ & 0.917  \\
\hline
144 & 07 Oct 2023 & 23:17:22~UT  & 23:25:00~UT  &  5.33$\times$10$^{-6}$   & 1.18$\times$10$^3$  & 532$^{\prime\prime}$, 62$^{\prime\prime}$ & 0.939  \\
\hline
145 & 19 Oct 2023 & 12:08:40~UT  & 12:10:47~UT  &  1.53$\times$10$^{-6}$   & 4.54$\times$10$^2$  & 820$^{\prime\prime}$, -351$^{\prime\prime}$ & 0.911  \\
\hline
146 & 02 Nov 2023 & 12:21:09~UT  & 12:22:32~UT  &  2.09$\times$10$^{-5}$   & 6.22$\times$10$^3$  & 461$^{\prime\prime}$, -359$^{\prime\prime}$ & 0.993  \\
\hline
147 & 05 Nov 2023 & 11:35:57~UT  & 11:43:06~UT  &  2.18$\times$10$^{-5}$   & 2.32$\times$10$^3$  & -755$^{\prime\prime}$, -192$^{\prime\prime}$ & 0.989  \\
\hline
148 & 05 Nov 2023 & 14:31:06~UT  & 14:32:49~UT  &  1.91$\times$10$^{-5}$   & 5.25$\times$10$^3$  & -745$^{\prime\prime}$, -193$^{\prime\prime}$ & 0.945  \\
\hline
149 & 11 Nov 2023 & 03:55:28~UT  & 03:59:20~UT  &  7.65$\times$10$^{-6}$   & 1.56$\times$10$^3$  & 566$^{\prime\prime}$, 90$^{\prime\prime}$ & 0.985  \\
\hline
\end{longtable}
\end{center}

\end{document}